\documentstyle[graphicx]{elsart}

\begin{document}
\begin{frontmatter}

\title{
Cryogenic measurement of the optical absorption coefficient in sapphire
crystals at 1.064\,$\rm{\mu} \rm{m}$
for the Large-scale Cryogenic Gravitational wave Telescope
}

\author{Takayuki Tomaru\thanksref{tom}}
\author{,Takashi Uchiyama\thanksref{uchi}}
\author{, Daisuke Tatsumi\thanksref{tatsu}}
\author{, }
\author{Shinji Miyoki, Masatake Ohashi, Kazuaki Kuroda}
\address{
Institute for Cosmic Ray Research (ICRR), The University of Tokyo, 5-1-5,
Kashiwanoha, Kashiwa, Chiba, 277-8582, Japan
}

\author{
Toshikazu Suzuki, Akira Yamamoto, Takakazu Shintomi
}
\address{
High Energy Accelerator Research Organization (KEK), 1-1 Oho, Tsukuba,
Ibaraki, 305-0801, Japan
}

\thanks[tom]{Corresponding author.
E-mail address: tomaru@icrr.u-tokyo.ac.jp}
\thanks[uchi]{Present address: High Energy Accelerator Research
Organization (KEK), 1-1 Oho, Tsukuba, Ibaraki, 305-0801, Japan}
\thanks[tatsu]{Present address: National Astronomical Observatory (NAO),
Mitaka, Tokyo, 181-8588, Japan}

\begin{abstract}
We have applied laser calorimetry to the measurement of optical absorption in
mono-crystalline sapphire at cryogenic temperatures. Sapphire
is a promising candidate for the mirror substrates of the Large-scale
Cryogenic Gravitational wave Telescope.
The optical absorption coefficients of different sapphire samples at a
wavelength of
1.064\,$\rm{\mu} \rm{m}$ at 5\,K were found to average
90\,ppm/cm.
\end{abstract}


\end{frontmatter}

\section{Introduction}
Large
scale laser interferometers
such as TAMA\cite{tama}, LIGO\cite{ligo}, VIRGO\cite{virgo} and
GEO\cite{geo}, are being developed for the direct detection of
gravitational waves (GW).
TAMA has already reached a stage where high quality data can be taken for
several hours at a time\cite{data}.
However, much more sensitive detectors are planned
because the estimated GW event rate for coalescing neutron star binaries is
extremely low
even within the Virgo cluster (at a radius of 20\,Mpc, and the main
target of LIGO and
VIRGO).
The important limitations to the sensitivity of these interferometers
are seismic
vibration, thermal Brownian noise of mirrors and their
suspensions\cite{thnoise}, and photon shot noise.
Although fused silica is used in present interferometers as the main mirror
substrate, it is not the best material for advanced
GW interferometers,
due to concerns about thermal Brownian noise and thermal
lensing\cite{thlensa,thlensb} at very high optical power.
Another promising candidate material is mono-crystalline sapphire, but concerns
about thermo-elastic noise\cite{thelas1,thelas2} render sapphire
unsuitable if used at room temperature.

We have been developing a cryogenic mirror technique to be used in the
Large-scale Cryogenic Gravitational wave Telescope (LCGT)\cite{lcgt}.
Sapphire will be used due to its extremely high Q\cite{cryoQ}, large
thermal conductivity and small thermal expansion coefficient at
cryogenic temperature.
These characteristics drastically reduce the effects of thermal Brownian
noise, thermal lensing and thermo-elastic noise\cite{cryothelas}.
However, concerns have been raised about possible large optical absorption in
sapphire,
which would lead to increased thermal lensing.
Room temperature measurements of optical absorption in
sapphire reported by several groups exhibit a wide spread of values
from 3\,ppm/cm to
140\,ppm/cm,
even where measurements were made on the same sample\cite{absB,absL}.
In these measurements, the photothermal technique\cite{method,intcal} was
used, which is an indirect method.

As a fundamental study towards the development of LCGT, we measured
the optical absorption
coefficient in sapphire at cryogenic temperature using laser
calorimetry\cite{method,thcal}.
Laser calorimetry at cryogenic temperatures has merit as a measurement method;
\begin{enumerate}
\item Since it is a direct measurement, it doesn't rely on detailed
knowledge of other material parameters
such as specific heat, thermal conductivity and temperature coefficient of
refractive index.
\item Since the thermal radiation from samples is very small at cryogenic
temperature,
small absorbed laser power makes a relatively large temperature
increase, easily measured to high precision.
\item Since the temperature in the cryostat is very stable, this
measurement technique is very insensitive to changes in the surroundings.
\item Highly sensitive thermometers are available for the measurement of
cryogenic temperatures.
Carbon-Glass Resistance (CGR) thermometers were used in this measurement,
which have an accuracy of better than 1\,mK near liquid helium
temperatures.
\end{enumerate}

\section{The principle of measurement}

\begin{figure}[htb]
\begin{center}
\includegraphics[height=8cm]{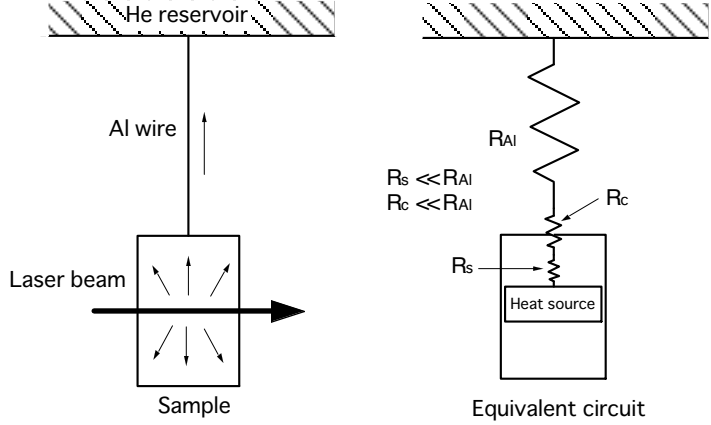}
\caption{Equivalent circuit of this measurement. $R_s$: Thermal resistance
of the sample, $R_c$: Contact resistance between the sample and the
aluminum wire, $R_{Al}$: Thermal resistance of the aluminum wire.
$R_{Al}$ was chosen to be much larger than $R_s$ and $R_c$. }
\label{eqcircuit}
\end{center}
\end{figure}

Changes are measured in the steady state
temperature of the sample for varying incident laser powers, after cooling
the sapphire sample to liquid helium temperature
using thermal conduction through an aluminum wire.
In short, this measurement is equivalent to measuring the thermal
resistance of the aluminum wire.
Figure \ref{eqcircuit} shows the equivalent circuit of this measurement.
Since the thermal resistance in the sapphire sample is much smaller than
one of the aluminum wire,
we can ignore the distribution of temperature within the sample.
We can formulate a thermal equation for the steady state temperature
$T(x)$ in the aluminum wire and
the two corresponding boundary conditions as follows:
\begin{eqnarray}
- \kappa \frac{d^2T(x)}{dx^2} = 0, \\
T(0) = T_0, \\
\kappa \frac{dT(L)}{dx} S = P,
\end{eqnarray}
where $\kappa$ is the thermal conductivity of the aluminum wire,
$T_0$ is the temperature of the end of the aluminum wire connected to the
helium reservoir
(equivalent to the initial temperature of the sample),
$L$ is the length of the aluminum wire,
$S$ is the cross sectional area of the aluminum wire,
and $P$ is the input heat power into the sample, which is equivalent to
the absorbed laser power.
The origin of the x-axis was chosen at the end of the aluminum wire.
Generally, thermal conductivity $\kappa$ depends on temperature,
however we can assume it to be constant because the change of the sample
temperature is at largest 100\,mK.
The error caused by this assumption is at most a few percent.
Integrating the above equations, the steady state temperature at the
sample $T(L)$ can be written as,
\begin{eqnarray}
T(L) = R_{Al}\cdot P +T_0, \\
R_{Al} \equiv \frac{L}{\kappa S}.
\end{eqnarray}
The steady state temperature at the sample is proportional to the input
power.
The thermal resistance $R_{Al}$ was determined using a heater that
produces known heat power.
Other errors concerning heat flux are also canceled by calibrating in this way.
After calibration, we can obtain the optical absorption coefficient $\alpha$,
\begin{equation}
\alpha = \frac{1}{l} \frac{P_{abs}}{P_{las}} = \frac{1}{l P_{las}} 
\frac{T(L) - T_0}{R_{Al}},
\end{equation}
where $P_{abs}$ is the laser power absorbed in the sample, $P_{las}$ is the
laser power injected into the sample
and $l$ is the length of the sample.

\begin{figure}[htb]
\begin{center}
\includegraphics[height=8cm]{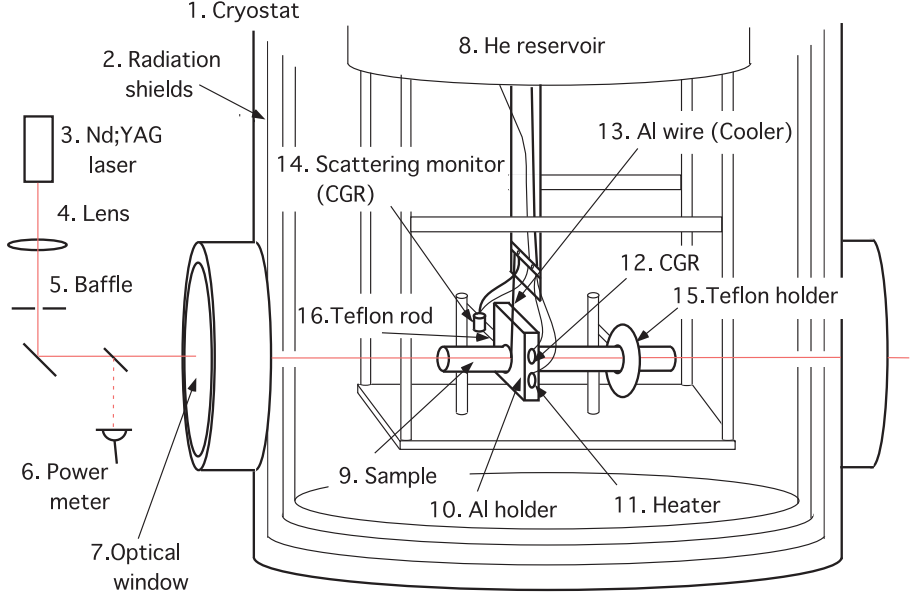}
\caption{The setup for the optical absorption measurement in the sapphire
crystals at cryogenic temperature.
1; Cryostat, 2; Radiation shields, 3; Nd:YAG laser, 4; Lens, 5; Baffle, 6;
Power meter, 7; Optical window,
8; Helium reservoir, 9; Sample, 10; Aluminum holder, 11; Manganin heater,
12; CGR thermometer, 13; Aluminum wire (Cooler), 14; Scattering monitor
(CGR thermometer), 15; Teflon holder, 16; Teflon rod.}
\label{setup1}
\end{center}
\end{figure}

Figure \ref{setup1} shows the experimental setup.
A 1.064\,$\rm{\mu} \rm{m}$ Nd:YAG laser was used in this measurement.
This laser has 700\,mW output power with a power stability of 0.1\,\%.
Injected laser power was measured by a power meter with an accuracy of
3\,\% and
net laser power in the sample was calculated considering multi-reflection
within the sample\cite{multiref}.
The sapphire sample was held in a pure aluminum mounting, itself mounted
on Teflon rods.
An aluminum wire was tightened between the holder and the sample in a
crush joint, and thermally connected to the helium reservoir.
A CGR thermometer and a manganin heater were mounted in the holder.
The thermometer and heater wires were both manganin and
superconducting.
The diameter and length were optimized to be able to ignore both the
production and the conduction of heat.
We could measure a small thermal contact resistance between the sample and the
aluminum holder,
and we corrected for this
after post-experiment measurement of the crushed area of the aluminum
wire. Since some reports mention that the sensitivity of laser calorimetry is
limited by the heat produced by light scattered
from the sample to the thermometer\cite{method}, we suspended another
CGR thermometer as a scattering monitor near the sample to further
investigate.

\section{Result}
We measured two mono-crystalline sapphire samples,
both manufactured by Crystal Systems Inc. using the Heat Exchange
Method\cite{hem}.
The grades of these samples were specified as "CSI white high purity" and
"Hemlite", respectively.
These sapphire grades are characterized by the homogeneity of
refractive index.
Typical homogeneity of the refractive index for CSI white is $1\times
10^{-6}$ and
that for Hemlite is $3\times 10^{-6}$\cite{hem}.
The CSI white sample was 10\,mm in diameter and 150\,mm in length
(cylinder axis was parallel to the c-axis).
The Hemlite sample was 100\,mm in diameter and 60\,mm in thickness
(again parallel to the c-axis).
All surfaces of the samples were optically polished.
Though these samples had been annealed during the process of crystal growth,
they were not re-annealed after polishing.
We measured at three spatially different points on each sample to confirm
that our measurement was not affected by the heat produced by
surface dust or defects, and to examine whether there was
any inhomogeneity of absorption.
Measurements were repeated more at least twice at each point.
The samples were cooled to 5\,K and
the temperature rise due to absorption of laser power was at most 100\,mK.
\begin{figure}[hbt]
\begin{center}
\includegraphics[height=5.5cm]{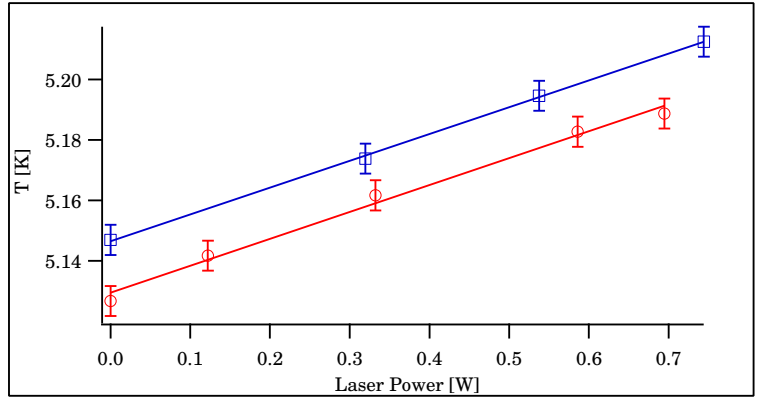}
\caption{The steady state temperatures at point 1 in the CSI white sample
risen by some injecting laser powers.
Measurements were done twice in this point.
The open circles show the first measurement and open squares show the next
measurement.}
\label{abs}
\end{center}
\end{figure}
Figure \ref{abs} shows the steady state temperature at the first
measurement point in the CSI white sample.
Measurements were done twice at this point.
The error bars were derived from the maximum fluctuation of electrical
output from the thermometer, which was $\pm$ 0.3\,\%,
corresponding to $\pm$ 5\,mK temperature error.
A small drift of the initial temperature of these two measurements was
caused by a change of a depth of liquid helium
in the reservoir, however this drift is much slower than the measurement
time of absorption.
We are only interested in the derivatives, which are then compared to the
calibration made using the heater.

\begin{table}[htdp]
\caption{The results of the optical absorption coefficients in sapphire
samples at 1.064\,$\rm{\mu} \rm{m}$ at 5\,K.}
\begin{center}
\begin{tabular}{|c|c|c|}
\hline
Point & CSI white high purity [ ppm/cm ] & Hemlite [ ppm/cm ] \\
\hline
1 & $93 \pm 9 $ & $99 \pm 13 $ \\
\hline
2 & $88 \pm 12$ & $90 \pm 10 $ \\
\hline
3 & $93 \pm 10$ & $90 \pm 10$ \\
\hline
\end{tabular}
\end{center}
\label{res}
\end{table}%
Table \ref{res} shows the measured optical absorption coefficient at
each point.
The optical absorption coefficients in the CSI white sample ranged from
88\,ppm/cm to 93\,ppm/cm.
The optical absorption coefficients in the Hemlite sample ranged from
90\,ppm/cm to 99\,ppm/cm.
The errors were about $\pm$ 10\,ppm/cm for all measurements.
No heat production by light scattering from the sample was observed at the
scattering monitor (CGR).
We did not find a large difference in the optical absorption between our
cryogenic result and a previous report at room temperature\cite{absL}.

\section{Conclusion}
We measured optical absorption in two sapphire samples, which were
manufactured by Crystal Systems Inc., at 1.064\,$\rm{\mu} \rm{m}$
wavelength at 5\,K, using a 700\,mW laser.
The optical absorption coefficients for the CSI white sample ranged from
88\,ppm/cm to 93\,ppm/cm.
The optical absorption coefficients for the Hemlite sample ranged from
90\,ppm/cm to 99\,ppm/cm.
In both samples, the measurement errors were about $\pm$ 10\,ppm/cm.

Our studies of thermal lensing have already confirmed that thermal
lensing will be negligible in interferometric GW detectors which use 
cryogenic sapphire mirrors. This research will be fully reported in 
an upcoming paper. The only remaining problem caused by sapphire's 
large optical absorption is mirror cooling. In the case of LCGT, 
which proposes a 10\,cm mirror length and 2.5\,kW of laser power 
passing through each of the near
mirrors, a heat power of 2\,W is deposited by direct optical 
absorption (assuming 90\,ppm/cm). From previous studies of cryogenic 
sapphire mirrors\cite{cooling,UDron}, the optical absorption must be reduced by 
one order of magnitude to allow LCGT's sapphire mirrors to operate at 
a temperature to 30\,K. This calculation assumes a 1\,mm diameter and 
250\,mm length for each of the the four sapphire fibers used as heat 
conductors. The integrated thermal conductivity of the sapphire 
fibers between 4.2\,K and 30\,K was taken to be $3\times 10^4 \mathrm{W/m}$\cite{UDron}.

The sources of the optical absorption are suspected to be impurities or
lattice defects.
We have confirmed the presence of $\mathbf{Ti^{3+}}$, $\mathbf{Cr^{3+}}$
and other unidentified impurities in these samples.
However, we have not yet identified the true sources of optical absorption
at 1.064\,$\rm{\mu} \rm{m}$.
This problem will be addressed in a future study.

The cryogenic measurement of optical absorption established in this study
can be used in the development of optical components for the advanced
interferometric gravitational wave detectors.

\ack
This study was supported by the Joint Research and Development Program of
KEK
and by a grant-in-aid from the Japanese Ministry of Education, Science, Sports
and Culture.
We thank Dr. C. T. Taylor for useful advice in the preparation of this
manuscript.

\end{document}